# TBA SCHEME WITH ION/PROTON DRIVING BEAM

E.G. Bessonov, FIAN LPI, Moscow, Russia
A.A. Mikhailichenko[#], Cornell University, LEPP, Ithaca, NY 14853, U.S.A.

*Abstract*

We are considering a two-beam accelerator (TBA) scheme with ion or proton beam as a driver. By comparison of the proposed scheme and the one with electron driver, we concluded, that TBA with ion/proton driver beam looks preferable. Existence of big proton accelerators in a few laboratories gives a boost for reconsideration of the baseline for post-LHC era. These Labs are FERMILAB, BNL, CERN and IHEP at Protvino, Moscow region. Protvino could emerge as one advantageous location and get stimulus for recovering the 600*GeV*-proton synchrotron in the existing ~20*km*-long tunnel. This synchrotron was planned as a booster for 3x3*TeV* storage ring.

## OVERVIEW

Many authors have developed TBA during the last decades [1]. CLIC is the mostly advanced representative of this kind [2]. The CLIC team does not give up even after International Technology Recommendation Panel made their decision in a favour of SC technology in August 2004. This is a good indication that some positive aspects are present in this idea. Obvious difficulty of TBA scheme associated with generation of electron driving beam (which forced recent change of CLIC operational frequency, by the way). To be useful for excitation of accelerating structure, the driving beam should have maximal content of spectral component of the driving current at the operational frequency. To some extent, TBA scheme with electron beam as a driver uses low impedance beam for transferring its energy to a high impedance one.

On the other hand, an idea of energy accumulation in a beam circulating in a storage ring and further usage of it for excitation of RF structure is an old one, discussed by G.I.Budker [3]. Later the idea to use the proton beam for excitation of the accelerating structure of electron linac was revealed in [4]. Here the proton beam excites *the same structure*, which is used for acceleration of electrons (or positrons). Naturally, this narrows the freedom of optimization of RF generation and further transferring it to the accelerating beam, as the transfer structure should take only a small fraction of power from the drive beam, while the accelerating structure should deliver as much power to the beam as possible. Usage of different structures for extraction of energy and for acceleration, linked together by the waveguides solves this problem.

The proton/ion drive beam is more advantageous, than the electron one is as follows: first advantage of the ion/proton beams is associated with their much lower emittance. These beams (or plans to have them) already

---
[#]aam10@cornell.edu

exist–that is another advantage. Other positive moment associated with lower gamma factor $\gamma$ for the same energy $\sim mc^2\gamma$. Lower gamma factor allows easier manipulation of the beam in a longitudinal phase space, $\sim 1/\gamma^2$. High stored energy in the proton beam (up to few *MJ*) is more than enough for excitation of RF structure. For fixed radius of accelerator the intensity of synchrotron radiation $\sim \gamma^2$, which excludes the losses associated with SR for protons. These losses prohibit usage of electron beam with high-energy as a driver. On the other hand, the energy of proton beam is high, so the ratio of impedances of the driving beam to the main beam is closer to unity for the proton/ion driver. The longitudinal component of the transport current is the only important parameter in a process of RF generation in a transfer structure. Lower $\gamma$ makes bunching with chicane easier $\sim 1/\gamma^2$ and decreases the longitudinal mass $\sim m\gamma^3$.

Basically, we raise the question for revision of TBA scheme in a favour of proton/ion driver beam. We shall use 30GHz-CLIC parameters as the reference ones [11]. All components developed for this project can be used for our scheme, delivering substantial savings.

## PRINCIPAL SCHEME

Principal scheme of TBA driven by the Ion/Proton beams is represented in Fig.1.

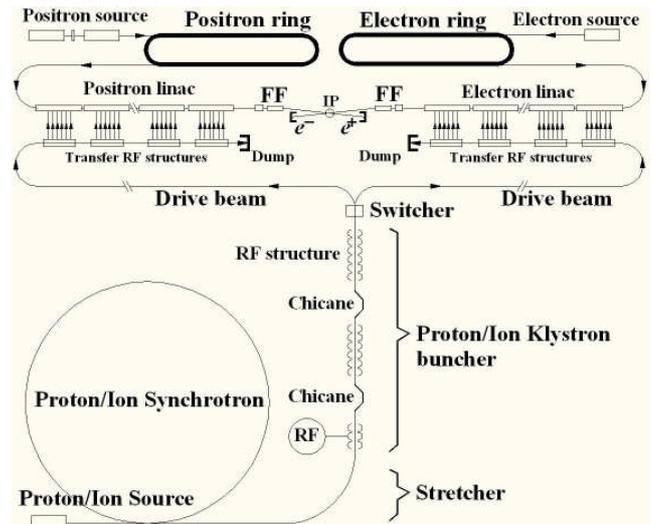

Figure 1: Principal scheme of the Complex proposed. FF stands for the Final Focus, IP-interaction point. RF stands for the RF generator feeding the first bunching structure.

Positron source with undulator could be easily introduced here in the same style as the one for ILC [5]. We consider the possibility of stacking polarized positrons obtained by conversion of polarized electrons in a thin target as well [6].

Below we describe the key elements of this scheme.

*Proton/Ion klystron buncher*

Operation of the Proton/Ion klystron buncher is basically the same as a buncher in usual klystron. As the beam is relativistic, the drift space replaced by the chicanes with a big value of $R_{56} = \kappa = \gamma(\partial l/\partial \gamma)$. This is actually a multi-cavity klystron, where the accelerating structures serve as cavities. The ratio of DC to AC current could reach a factor of 2 with a multi-frequency bunching.

Details of klystron operation one can find anywhere, we will underline here some key points. Distribution of particles (protons/ions) in the longitudinal phase space $\{\Delta E, \vartheta\}$ can be characterized by a function of these two variables $f(\Delta E, \vartheta)$, where the energy deviation $\Delta E$ and the phase $\vartheta = \omega t$ are canonically conjugated; $\omega$ stands for the frequency, $t$ –for the time. These variables are so called characteristics of the equations of the longitudinal motion. Motion in a phase space could be described by the generation function, which is action, cinematically transferring these variables from one time slice to another one $(\Delta E, \vartheta)_1 \to (\Delta E, \vartheta)_2$ [7],[8]. Function $f(\Delta E, \vartheta)$ stays invariant under these transformations. Therefore, for description of the phase distribution in actual point, one needs cinematically transform the coordinates of individual particle to this point. In a good assumption, the distribution function could be factorized as $f(\Delta E, \vartheta) = f(\Delta E) \cdot f(\vartheta)$, and the Fourier component of current at the second point is

$$i_m = \iint f(\Delta E) f(\vartheta_2) e^{-im\vartheta_2} v d\vartheta_2 d\Delta E \qquad (1)$$

where $v$ is a speed of particle. Initially, $f = 1/2\pi$, and $f(\Delta E) = \frac{1}{\sqrt{2\pi}\sigma_E} exp(-\Delta E^2/2\sigma_E^2)$, where $\sigma_E$ stands for the local energy spread in the beam, variable $\vartheta_2 = \omega t_2$ is an actual phase at the second point. Integral (1) can be transferred as the following

$$i_m = \frac{1}{2\pi} I_0 \int |\partial \vartheta/\partial \vartheta_2| e^{-im\vartheta_2} f(\Delta E) d\vartheta_2 d\Delta E =$$
$$= \frac{1}{2\pi\sqrt{2\pi}\sigma_E} I_0 \iint e^{-im\vartheta_2} exp(-\Delta E^2/2\sigma_E^2) d\vartheta_2 d\Delta E \qquad (2)$$

where $I_0$ is the local current at the start point, $(\Delta E/E)_\sigma$ is the relative energy deviation. By introduction $X_{12} = k\kappa(eU_{eff}/E)$, $k = \omega/v$, $\vartheta_{12} = \omega l/v = kl$, $U_{eff}$ is effective voltage of the first cavity (structure) one can rewrite (2)

$$i_m = \frac{1}{2\pi} I_0 e^{i\vartheta_{12}} \int exp(im(\vartheta_1 - X_{12}Cos\vartheta_1)) d\vartheta_1 \times$$
$$\times \frac{1}{\sqrt{2\pi}\sigma_E} \int exp(ik\kappa(\Delta E/E) - (\Delta E)^2/2\sigma_E^2) \cdot d\Delta E \qquad (3)$$

By integration one can find finally

$$i_m/I_0 = J_m(X_m) exp(-\tfrac{1}{2}(mk\kappa\sigma_E/E)^2) \qquad (4)$$

where $X_m = mX_{12}$, $J_m$ is a Bessel function of m-th order.

The term with exponent reflects the reduction of AC current due to the energy spread in the bunch. The first harmonic, $m=1$ is the subject of interest; $J_1(X_1)$ has a maximum at $X_1 = 1.82$, $J_1 = 0.57$. For $\sigma_E/E \cong 0.001$, $E \approx 600 GeV$, $\lambda \sim 1 cm$; debunching due to energy spread being low so the efficiency of this system will be ~57%. If the beam comes to the second cavity at the phase $\omega t_2$, then the modulation provided by the second cavity becomes

$$|\Delta E/E|_2 Cos(\omega t_1 + \vartheta_{12} + X_{12} \cdot Cos(\omega t_1)) \qquad (5)$$

It is clear from (5) that many frequencies are generated, due to appearance of the cosine function as an argument of a cosine. The main result from this is the presence of second harmonic in the phase modulation due to *cascade* bunching. In fact, the second harmonic can be obtained in a dedicated RF structure (cavity) with second harmonic feed by the additional RF generator. As a result, the level of the first harmonic becomes 28% higher than in a case with single-stage bunching, brining efficiency to ~75%. The voltage in a bunching cavity required for obtaining the bunching coefficient ~1 defined from the equity

$$1 \cong X \cong k\kappa \cdot (eU_{eff}/E) \cong (2\pi/\lambda) \cdot \kappa \cdot (eU_{eff}/E). \qquad (6)$$

The last expression gives the energy $eU_{eff}$ ~3 *MeV* only. If we suggest that the length of RF structure is 3 *m*, then the electric field strength should be $1MeV/m$, which could easily be realized (even with Superconducting RF). Practically there is no beam loading here.

The 600*GeV* beam can transfer its energy with ~75% efficiency to RF. If we suggest that the proton bunch population is $2 \cdot 10^{11}$ (see Table 1), then the energy of electron bunch with population ~$10^{10}$ can reach ~1*TeV* with 12% beam loading.

*Chicane*

Instead of straight section like in an ordinary klystron, used here is a chicane, with significant parameter

$$\kappa = \gamma(\partial l/\partial \gamma) = \int_0^s (D(s)/\rho(s)) \cdot ds, \qquad (7)$$

where $D(s)$ is a dispersion, $\rho(s)$ is a local bending radius in the magnets. For a three-magnet scheme [7]

$$\kappa = \frac{(2\pi)^2}{48}(1 + K_{bend}^2) \cdot s/\gamma^2 \qquad (8)$$

where $K_{bend} \approx eH_0 s/2\pi m_p c^2$, $H_0$ is the bending field in the magnet, the *s* is a total length of the chicane. For example, if *s*=10*m*, $H_\perp = 2T$, then $K_{bend} \cong 2$ $\gamma \cong 600$, and $\kappa \cong 0.01 m$. The lengthening due to the natural energy spread in the beam will be $\Delta l \cong \lambda_{ac}/100$. The lengthening due to finite emittance can be made small as well.

*Stretcher*

At the stretcher, Fig.1, value of $R_{56}$ allows reduction of local energy spread by controllable enlargement of the bunch length. The beam structure can be any, even as long the circumference of synchrotron, if RF turned on at the top. Each bunch allowed expansion of its length while the local energy spread is decreasing in the same proportion as the length growth.

*Switcher*

The switcher serves for re-direction of bunch-trains to the electron/positron wings of collider. It is basically, a fast kicker. Switcher located in a slightly asymmetric position with respect to the IP, for proper phasing.

*Electron/Positron rings*

Electron/Positron rings use the same ideology as the ring suggested in [9]. They have long straight sections filled with the wigglers circled with the multi-magnet bends. Wigglers have linear piecewise field dependence for reduction of nonlinearities [10].

*Proton/Ion Synchrotron*

In UNK, Protvino, the booster synchrotron was planned for installation in the same main tunnel. That is why it has such a big circumference, very much desirable for our purposes, however. Parameters of existing synchrotrons represented in a Table 1. One other possibility is to fit a newly built synchrotron in existing main tunnel of appropriate Laboratory (IHEP-22.7*km*; CERN-26.67*km*; FERMILAB -6.28*km*; BNL-3.834*km*).

*Stretcher*

Typical bunch length in a proton accelerator is~1nsec (see Table 1). It is interesting that for 30*GHz*-CLIC scheme, the driving bunch train of 22 bunches was planned to be ~0.72 nsec long [11]. With the stretcher, the length of the train can be adjusted to any necessary value. Many designs can be found elsewhere, (see section *Chicane* above).

Table 1. Parameters of proton/ion synchrotrons

| Laboratory/location | IHEP/Protvino | CERN/Geneva | FERMILAB/Batavia | BNL/Brookhaven |
|---|---|---|---|---|
| Installation | UNK-600 | SPS | Main Injector | AGS |
| Energy | 600 $GeV$ | 450 $GeV$ | 150 $GeV$ | 24.5(28) $GeV$ |
| Circumference | 20.77$km$ | 6.9$km$ | 3.319 $km$ | 0.807 $km$ |
| Acceleration/flattop | 11/20 s | 4/3 s | 2.5/<0.1 s | (0.1-2)/0.03 s |
| $\Delta E/E$ (4$\sigma$) | $2\cdot 10^{-3}$ | $1.16\cdot 10^{-3}$ | $6\cdot 10^{-3}$ | $2\cdot 10^{-3}$ |
| Population/bunch | $4.0\cdot 10^{11}$ | $1.15\cdot 10^{11}$ | $(0.6-1.2)\cdot 10^{11}$ | $4.0\cdot 10^{13}$ |
| Number of bunches | 30x12 | 72x4 (4200) | 498/588 (max) | 8(24) |
| Bunch length/$c$ | - | 1.8 $ns$ | 10 $ns$ | 1 $ns$ |
| Bunch spacing/$c$ | 160 $ns$ | 25 $ns$ | 19 $ns$ | 224(336) $ns$ |
| RF Voltage max | 8 $MV$ | 7 $MV$ | 4 $MV$ | 147 $kV$ |
| Emittance transv. | 30 $\mu m$(norm) | 3.5$\mu m$ (450$GeV$) | 40 $\mu m$ (norm) | 50 $\mu m$(norm) |
| Emittance longitud. | 1 $eV$-s | 1 $eV$-s | 0.2 $eV$-s | 0.3 $eV$-s |

## SUMMARY

Existences of big proton/ion synchrotrons in few laboratories give a new boost for reconsidering the baseline for the post LHC era. These Labs are FERMILAB, BNL, CERN and IHEP at Protvino, Moscow region. Protvino could emerge as the most advantageous place for recovery of the developed proton synchrotron in existing ~ 20 km-long tunnel. This synchrotron was planned to be a booster for 3x3*TeV* UNK complex [12].